\documentclass[letterpaper, 10 pt, conference]{ieeeconf}  

\IEEEoverridecommandlockouts                              

\usepackage{amsmath}
\usepackage{amssymb}
\usepackage{tikz}
\usepackage{pgfplots}
\usepackage{graphicx}
\usepackage{caption}
\usepackage{subcaption}
\usepackage{xcolor}
\allowdisplaybreaks  
\usepackage{cite}
\usepackage{algorithmicx}
\usepackage{algpseudocode}
\usepackage{algorithm}

\allowdisplaybreaks

\newtheorem{definition}{Definition} 
\newtheorem{theorem}{Theorem} 
\newtheorem{problem}{Problem} 
\newtheorem{remark}{Remark}
\newtheorem{lemma}{Lemma}
\newtheorem{corollary}{Corollary}

\title{\LARGE \bf
Efficient Automata-based Planning and Control\\ under Spatio-Temporal Logic Specifications
}

\author{Lars Lindemann and Dimos V. Dimarogonas
\thanks{This work was supported in part by the Swedish Research Council (VR), the European Research Council (ERC), the Swedish Foundation for Strategic Research (SSF),  the EU H2020 Co4Robots project, and the Knut and Alice Wallenberg Foundation (KAW).}
\thanks{The authors are with the Division of Decision and Control Systems, School of Electrical Engineering and Computer Science, KTH Royal Institute of Technology, 100 44 Stockholm, Sweden. {\tt\small llindem@kth.se (L. Lindemann), dimos@kth.se (D.V. Dimarogonas)}}%
}

\begin{document}

\maketitle
\thispagestyle{empty}
\pagestyle{empty}

\begin{abstract}
The use of spatio-temporal logics in control is motivated by the need to impose complex spatial and temporal behavior on dynamical systems, and to control these systems accordingly. Synthesizing correct-by-design control laws is a challenging task resulting in computationally demanding methods. We consider efficient automata-based planning for continuous-time systems under signal interval temporal logic specifications, an expressive fragment of signal temporal logic. The planning is based on recent results for automata-based verification of metric interval temporal logic. A timed signal transducer is obtained accepting all Boolean signals that satisfy a metric interval temporal logic specification, which is abstracted from the signal interval temporal logic specification at hand. This transducer is modified to account for the spatial properties of the signal interval temporal logic specification, characterizing all real-valued signals that satisfy this specification. Using logic-based feedback control laws, such as the ones we have presented in earlier works, we then provide an abstraction of the system that, in a suitable way, aligns with the modified timed signal transducer. This  allows to avoid the state space explosion that is typically induced by forming a product automaton between an abstraction of the system and the specification.
\end{abstract}

\section{Introduction}
\label{sec:introduction}

The control of dynamical systems under complex temporal logic specifications has lately received increasing attention. One can distinguish between temporal logics that allow to express qualitative, e.g.,  linear temporal logic (LTL) \cite{pnueli1977temporal}, and quantitative, e.g., metric interval temporal logic (MITL) \cite{alur1996benefits}, temporal properties. An MITL specification can be translated into a language equivalent timed automaton \cite{alur1996benefits}. If the accepted language of this automaton is not empty, the MITL specification is satisfiable. Emptiness can be checked by abstracting the timed automaton into its untimed region automaton \cite{alur1994theory}. There exists no tool to algorithmically translate an MITL specification, interpreted over continuous-time semantics, into its language equivalent timed automaton. For point-wise semantics, such a tool has been presented in \cite{brihaye2017m}. Point-wise semantics, however, do not guarantee the satisfaction of the MITL specification in continuous time. The procedure of \cite{alur1996benefits}, for continuous-time semantics, is complex and rather of theoretical nature. The results from \cite{maler2006mitl,ferrere2019real} are more intuitive and present a compositional way to construct a timed signal transducer for an MITL specification. More recently, spatio-temporal logics have been considered that further allow to reason about spatial properties. Such spatio-temporal logics are signal temporal logic (STL) \cite{maler2004monitoring} or a variant of MITL where propositions are associated with observation maps \cite{fainekos2009robustness}. The richness and complexity of the chosen temporal logic increases by going from qualitative to quantitative temporal properties as well as by going from non-spatial to spatial properties.

Classical control theoretical tools, which deal with invariance and stability of dynamical systems, are not rich enough to solely deal with the control problem at hand. Hence, automata-based tools have been used to divide a specification into subtasks that can be achieved sequentially by low-level feedback control laws. There exist numerous approaches for LTL \cite{fainekos2009temporal,kress2009temporal,belta2007symbolic} and for MITL \cite{zhou2016timed,nikou2016cooperative,fu2015computational,fernando_ecc,liu2014switching,verginis2019reconfigurable}.   The idea is to abstract the system into an automaton and to form a product automaton with an automaton representing the LTL/MITL specification. This procedure is subject to a computational blowup due to an exponential explosion in the resulting state space. Spatio-temporal logics have not leveraged automata-based results. Thus far, STL and the associated robust semantics have been used for the full STL fragment and only for discrete-time systems resulting in computationally demanding mixed integer linear programs \cite{raman1}. Other approaches have  maximized the robust semantics in optimization-based frameworks, resulting again in computationally expensive methods \cite{pant2018fly,mehdipour}, prone to get stuck in local minima. For continuous-time systems and fragments of STL, robust and computationally-efficient time-varying feedback control laws have been presented in \cite{lindemann2018control,lindemann2019decentralized}.  

We consider continuous-time systems under spatio-temporal logic specifications expressed in signal interval temporal logic (SITL), an expressive STL fragment where temporal operators can not be constrained by singular intervals. We remark that SITL is a more expressive fragment than the fragments of STL that have been considered in \cite{lindemann2018control,lindemann2019decentralized}. The SITL specification at hand is first abstracted into an MITL specification that is translated into its language equivalent timed signal transducer \cite{ferrere2019real}. This transducer is  modified  to account for the error induced by considering propositions (MITL) instead of predicates (SITL).  The modified timed signal transducer characterizes all real-valued signals that satisfy the SITL specification and it can hence be checked whether or not the specification is satisfiable. To the best of our knowledge, this is the first decidability result for STL interpreted over continuous-time semantics. We then use logic-based feedback control laws that can achieve finite-time reachability and invariance, such as for instance presented in \cite{lindemann2018control,lindemann2019decentralized},  to define a timed abstraction of the system. This abstraction aligns, in a suitable way, with the modified timed signal transducer. In particular, this abstraction considers transitions between boolean combinations of predicates instead of transitions between cells of the continuous state space. In this way, an explosion of the state space in the product automaton between the abstraction and the timed signal transducer can be avoided. This product typically induces $\mathcal{O}(mn)$ states where $m$ and $n$ are the number of states in abstraction and specification automaton, respectively, while our approach works directly on an automaton with $n$ or less states. The main contribution is hence an efficient planning and control framework for continuous-time systems under spatio-temporal logic specifications.

Sec. \ref{sec:backgound} presents preliminaries and problem formulation. Our proposed problem solution is stated in Sec. \ref{sec:strategy}. Simulations and conclusions are given in Sec. \ref{sec:simulations} and Sec. \ref{sec:conclusion}.

\section{Preliminaries and Problem Formulation}
\label{sec:backgound}

True and false are $\top$ and $\bot$ with $\mathbb{B}:=\{\top,\bot\}$; $\mathbb{R}$, $\mathbb{Q}$, and $\mathbb{N}$ are the real, rational, and natural numbers, respectively, while $\mathbb{R}_{\ge 0}$ and $\mathbb{Q}_{\ge 0}$ denote their respective nonnegative subsets; $\mathbb{R}_{> 0}$  denotes the positive real numbers.

\subsection{Real-time Temporal Logics}
\label{sec:STL_back}

Let $P$ be a set of propositions. Metric interval temporal logic (MITL) \cite{alur1996benefits} is based on propositions $p\in P$ as well as Boolean and temporal operators. The syntax is given by 
\begin{align*}
\varphi \; ::= \; \top \; | \; p \; | \; \neg \varphi \; | \; \varphi' \wedge \varphi'' \; | \; \varphi'  U_I \varphi''\;
\end{align*}
where $\varphi$, $\varphi'$, and $\varphi''$ are MITL formulas, $\neg$ and $\wedge$ denote negation and conjunction, respectively, and $U_I$ is the until operator with $I\subseteq \mathbb{Q}_{\ge 0}$ and $I$ not being a singleton. We define $\varphi' \vee \varphi'':=\neg(\neg\varphi' \wedge \neg\varphi'')$ (disjunction), $F_I\varphi:=\top U_I \varphi$ (eventually operator), and $G_I\varphi:=\neg F_I\neg \varphi$ (always operator). Let $\boldsymbol{d}:\mathbb{R}_{\ge 0}\to \mathbb{B}^{|P|}$ be a Boolean signal corresponding to truth values of the propositions in $P$ over time. Define also the projection of $\boldsymbol{d}$ onto  $p\in P$ as proj$_p(\boldsymbol{d}):\mathbb{R}_{\ge 0}\to \mathbb{B}$. The expression $(\boldsymbol{d},t)\models \varphi$ indicates that the signal $\boldsymbol{d}$ satisfies an MITL formula $\varphi$ at time $t$. The continuous-time semantics of an MITL formula \cite[Sec. 4]{ferrere2019real} are then defined as $(\boldsymbol{d},t)\models p$ iff proj$_p(\boldsymbol{d}(t))=\top$, $(\boldsymbol{d},t)\models \neg \varphi$ iff $(\boldsymbol{d},t)\not\models \varphi$, $(\boldsymbol{d},t)\models \varphi'\wedge\varphi''$ iff $(\boldsymbol{d},t)\models \varphi'$ and $(\boldsymbol{d},t)\models \varphi''$, and $(\boldsymbol{d},t)\models \varphi'U_I\varphi''$ iff $\exists t'' \in t+I$, $(\boldsymbol{d},t'')\models \varphi''$ and $\forall t' \in (t,t'')$, $(\boldsymbol{d},t')\models \varphi'$ where $t+I$ intuitively denotes an interval. 

We further define signal interval temporal logic (SITL), a fragment of signal temporal logic (STL) \cite{maler2004monitoring}, as a simple yet expressive spatio-temporal logic by excluding, similar to MITL, singular time intervals in the temporal operators. SITL considers, instead of propositions, predicates $\mu\in M$  where $M$ denotes a set of predicates. The truth value of $\mu$ is determined by a predicate function $h:\mathbb{R}^n\to\mathbb{R}$ so that, for $\boldsymbol{\zeta}\in\mathbb{R}^n$, $\boldsymbol{\zeta}\models \mu$ iff $h(\boldsymbol{\zeta})\ge 0$. An SITL formula is then an MITL formula over predicates. The SITL syntax is hence the same as for MITL formulas, but with predicates $\mu$ instead of propositions $p$. Let $(\boldsymbol{x},t)\models \phi$ denote that the signal $\boldsymbol{x}:\mathbb{R}_{\ge 0}\to\mathbb{R}^n$ satisfies  $\phi$ at time $t$. Let $(\boldsymbol{x},t)\models \mu$ iff $h(\boldsymbol{x}(t))\ge 0$, while the semantics for Boolean and temporal operators are the same as for MITL. An SITL formula $\phi$ is satisfiable if there exists $\boldsymbol{x}\in\mathbb{R}^n$ such that $(\boldsymbol{x},0)\models \phi$.
  

The symbols $\varphi$ and $\phi$ are used to distinguish between MITL and SITL formulas, respectively. We will consider, in particular, an SITL formula $\phi$ that consists of the predicates $\mu_i\in M$ with $i\in\{1,\hdots,|M|\}$ and abstract  $\phi$, in a first step, into an MITL formula $\varphi$ as follows. Associate with each predicate $\mu_i\in M$ a proposition $p_i$ and let $P:=\{p_1,\hdots,p_{|M|}\}$. Let then $\varphi:=Pr(\phi)$ be an MITL formula that is obtained by replacing each predicate $\mu_i\in M$ in $\phi$ with a proposition $p_i\in P$, e.g., $\phi:=F_{I}(\mu_1 \wedge \mu_2)$ becomes $\varphi:=Pr(\phi)=F_{I}(p_1\wedge p_2)$. This way, spatial properties of $\phi$ are neglected in $\varphi$. Conversely, let $Pr^{-1}(\varphi)=Pr^{-1}(Pr(\phi))=\phi$ be obtained by replacing each proposition $p_i\in P$ in $\varphi$ with the corresponding predicate $\mu_i\in M$.

\subsection{MITL to Timed Signal Transducer}
\label{sec:mitl_to_ta}

An MITL formula $\varphi$ can be translated into a language equivalent timed signal transducer \cite{ferrere2019real} by means of a simple compositional procedure, as summarized next.  Let $\boldsymbol{c}:=\begin{bmatrix} c_1 &\hdots &c_O \end{bmatrix}^T\in\mathbb{R}_{\ge 0}^O$ be a vector of $O$ clock variables that obey the continuous dynamics $\dot{c}_o(t):=1$ with $c_o(0):=0$ for $o\in\{1,\hdots,O\}$. Discrete dynamics occur at instantaneous times in form of clock resets. Let $R:\mathbb{R}_{\ge 0}^O\to \mathbb{R}_{\ge 0}^O$ be a reset function such that $R(\boldsymbol{c})=\boldsymbol{c}'$ where either $c_o'=c_o$ or $c_o'=0$. With a slight abuse of notation, we also use $R(c_o)=c_o$ and $R(c_0)=0$. Clocks evolve with time when visiting a state of a timed signal transducer, while clocks may be reset during transitions between states. We further define clock constraints as Boolean combinations of conditions of the form $c_o\le k$ and $c_o\ge k$ for some $k\in\mathbb{Q}_{\ge 0}$. Let $\Phi(\boldsymbol{c})$ denote the set of all clock constraints over clock variables in $\boldsymbol{c}$. 
\begin{definition}[Timed Signal Transducer \cite{ferrere2019real}]
A timed signal transducer is a tuple $TST:=(S,s_0,\Lambda,\Gamma,\boldsymbol{c},\iota,\Delta,\lambda,\gamma, \mathcal{F})$ where $S$ is a finite set of locations, $s_0$ with $s_0\cap S=\emptyset$ is the initial state, $\Lambda$ and $\Gamma$ are a finite sets of input and output variables, respectively, $\iota:S\to\Phi(\boldsymbol{c})$ assigns clock constraints over $\boldsymbol{c}$ to each location, $\Delta$ is a transition relation so that $\delta=(s,g,R,s')\in\Delta$ indicates a transition from $s\in S\cup s_0$ to $s'\in S$ satisfying the guard constraint $g\subseteq \Phi(\boldsymbol{c})$ and resetting the clocks according to $R$; $\lambda:S\cup\Delta\to BC(\Lambda)$ and $\gamma:S\cup\Delta\to BC(\Gamma)$ are input and output labeling functions where $BC(\Lambda)$ and $BC(\Gamma)$ denote the sets of all Boolean combinations over $\Lambda$ and $\Gamma$, respectively, and $\mathcal{F}\subseteq 2^{S\cup \Delta}$ is a generalized B\"uchi acceptance condition.
\end{definition}

A run of a $TST$ over an input signal $\boldsymbol{d}:\mathbb{R}_{\ge 0}\to \mathbb{B}^{|\Lambda|}$ is an alternation of time and discrete steps resulting in an output signal $\boldsymbol{y}:\mathbb{R}_{\ge 0}\to \mathbb{B}^{|\Gamma|}$. A time step of duration $\tau\in\mathbb{R}_{>0}$ is denoted by $(s,\boldsymbol{c}(t))\xrightarrow{\tau}(s,\boldsymbol{c}(t)+\tau)$ with $\boldsymbol{d}(t+t')\models\lambda(s)$, $\boldsymbol{y}(t+t')\models\gamma(s)$, and $\boldsymbol{c}(t+t')\models \iota(s)$ for each $t'\in (0,\tau)$. A discrete step at time $t$ is denoted by $(s,\boldsymbol{c}(t))\xrightarrow{\delta}(s',R(\boldsymbol{c}(t)))$ for some transition $\delta=(s,g,R,s')\in\Delta$ such that $\boldsymbol{d}(t)\models\lambda(\delta)$, $\boldsymbol{y}(t)\models\gamma(\delta)$, and $\boldsymbol{c}(t)\models g$. Each run starts with a discrete step from the initial configuration $(s_0,\boldsymbol{c}(0))$. Formally, a run of a $TST$ over $\boldsymbol{d}$ is a sequence $(s_0,\boldsymbol{c}(0))\xrightarrow{\delta_0}(s_1,R_0(\boldsymbol{c}(0)))\xrightarrow{\tau_1}(s_1,R_0(\boldsymbol{c}(0))+\tau_1)\xrightarrow{\delta_1} \hdots$. Due to the alternation of time and discrete steps, the signals $\boldsymbol{d}(t)$ and $\boldsymbol{y}(t)$ may be a concatenation of sequences consisting of points and open intervals. Zeno signals are excluded by assumption \cite{ferrere2019real}. We associate a function $q:\mathbb{R}_{\ge 0}\to S\cup \Delta$ with a run as $q(0):=\delta_0$, $q(t)=s_1$ for all $t\in(0,\tau_1)$, $\hdots$; $\mathcal{F}$ is a generalized B\"uchi acceptance condition so that a run over $\boldsymbol{d}(t)$ is accepting if, for each $F\in\mathcal{F}$, $\text{inf}(q)\cap F\neq \emptyset$ where $\text{inf}(q)$ contains the states in $S$ that are visited, in $q$, for an unbounded time duration and transitions in $\Delta$  that are taken, in $q$, infinitely many times. We define the language of $TST$ to be $L(TST):=\{\boldsymbol{d}:\mathbb{R}_{\ge 0}\to \mathbb{R}^{|\Lambda|}|TST\text{ has an accepting run over } \boldsymbol{d}(t)\}$. 
\begin{definition}[Synchronous Product \cite{ferrere2019real}]
Given $TST_i:=(S_i,s_{0,i},\Lambda_i,\Gamma_i,\boldsymbol{c}_i,\iota_i,\Delta_i,\lambda_i,\gamma_i, \mathcal{F}_i)$ with $i\in\{1,2\}$, their synchronous product is  $TST:=(S,s_0,\Lambda,\Gamma,\boldsymbol{c},\iota,\Delta,\lambda,\gamma, \mathcal{F})$ with $S:=S_1 \times S_2$, $s_0:=s_{0,1}\times s_{0,2}$, $\Lambda:=\Lambda_1\cup \Lambda_2$, $\Gamma:=\Gamma_1\cup \Gamma_2$, $\boldsymbol{c}:=\begin{bmatrix}\boldsymbol{c}_1^T & \boldsymbol{c}_2^T \end{bmatrix}^T$, and  $\iota(s_1,s_2):=\iota_1(s_1)\wedge \iota(s_2)$. The transition relation $\Delta$ is defined as
\begin{itemize}
\item $((s_1,s_2),g,R,(s_1',s_2'))\in\Delta$ where $(s_1,g_1,R_1,s_1')\in\Delta_1$, $(s_2,g_2,R_2,s_2')\in\Delta_2$, $g:=g_1\wedge g_2$, and $R:=\begin{bmatrix}R_1^T & R_2^T\end{bmatrix}^T$ (simultaneous transitions),
\item $((s_1,s_2),g_1 \wedge \iota_2(s_2),R_1,(s_1',s_2))\in\Delta$ where $(s_1,g_1,R_1,s_1')\in\Delta_1$ (left-sided transitions),
\item $((s_1,s_2),\iota_1(s_1)\wedge g_2,R_2,(s_1,s_2'))\in\Delta$ where $(s_2,g_2,R_2,s_2')\in\Delta_2$ (right-sided transitions),
\end{itemize}
and the input labeling function defined as
\begin{itemize}
\item $\lambda(s_1,s_2):=\lambda_1(s_1)\wedge \lambda_2(s_2)$ (state labels),
\item $\lambda((s_1,s_2),g,R,(s_1',s_2')):=\lambda_1(s_1,g_1,R_1,s_1')\wedge \lambda_2(s_2,g_2,R_2,s_2')$ (simultaneous transitions), 
\item $\lambda((s_1,s_2),g_1\wedge\iota_2(s_2),R_1,(s_1',s_2)):=\lambda_1(s_1,g_1,R_1,s_1')\wedge\lambda_2(s_2)$ (left-sided transitions), 
\item $\lambda((s_1,s_2),\iota_1(s_1)\wedge g_2,R_2,(s_1,s_2')):=\lambda_1(s_1)\wedge\lambda_2(s_2,g_2,R_2,s_2')$ (right-sided transitions)
\end{itemize}
while the output labeling function is constructed the same way as the input labeling function. The B\"uchi acceptance condition is $\mathcal{F}:=\{\mathcal{F}_1\times (S_2\cup \Delta_2), (S_1\cup\Delta_1)\times \mathcal{F}_2\}$.
\end{definition}
\begin{definition}[Input-Output Composition \cite{ferrere2019real}]\label{def:i-o-comp}
Given $TST_i:=(S_i,s_{0,i},\Lambda_i,\Gamma_i,\boldsymbol{c}_i,\iota_i,\Delta_i,\lambda_i,\gamma_i, \mathcal{F}_i)$ with $i\in\{1,2\}$, the input-output composition where the output of $TST_1$ is the input of $TST_2$ is $TST:=(S,s_0,\Lambda,\Gamma,\boldsymbol{c},\iota,\Delta,\lambda,\gamma, \mathcal{F})$ with $S:=\{(s_1,s_2)\in S_1 \times S_2|\text{if } \boldsymbol{d}\models \gamma_1(s_1)\text{ implies } \boldsymbol{d}\models\lambda_2(s_2)\}$, $\Lambda:=\Lambda_1$, $\Gamma:= \Gamma_2$, and $s_0$, $\boldsymbol{c}$, $\iota$, and $\mathcal{F}$ as defined in the synchronous product. The transition relation $\Delta$ is defined as
\begin{itemize}
\item $((s_1,s_2),g,R,(s_1',s_2'))\in\Delta$ where $\delta_1:=(s_1,g_1,R_1,s_1')\in\Delta_1$, $\delta_2:=(s_2,g_2,R_2,s_2')\in\Delta_2$, $g=g_1\wedge g_2$, and $R=\begin{bmatrix}R_1^T & R_2^T\end{bmatrix}^T$ if $\boldsymbol{d}\models \gamma_1(\delta_1)$ implies $\boldsymbol{d}\models\lambda_2(\delta_2)$ (simultaneous transitions),
\item $((s_1,s_2),g_1 \wedge \iota_2(s_2),R_1,(s_1',s_2))\in\Delta$ where $\delta_1:=(s_1,g_1,R_1,s_1')\in\Delta_1$ if $\boldsymbol{d}\models \gamma_1(\delta_1)$ implies $\boldsymbol{d}\models\lambda_2(s_2)$ (left-sided transitions),
\item $((s_1,s_2),\iota_1(s_1)\wedge g_2,R_2,(s_1,s_2'))\in\Delta$ where $\delta_2:=(s_2,g_2,R_2,s_2')\in\Delta_2$ if $\boldsymbol{d}\models\gamma_1(s_1)$ implies $\boldsymbol{d}\models\lambda_2(\delta_2)$ (right-sided transitions),
\end{itemize}
and the input and output labeling functions are defined as
\begin{itemize}
\item $\lambda(s_1,s_2):=\lambda_1(s_1)$ and $\gamma(s_1,s_2):=\gamma_2(s_2)$ (state labels),
\item  $\lambda((s_1,s_2),g,R,(s_1',s_2')):=\lambda_1(s_1,g_1,R_1,s_1')$ and $\gamma((s_1,s_2),g,R,(s_1',s_2')):=\gamma_2(s_2,g_2,R_2,s_2')$ (simultaneous transitions),
\item  $\lambda((s_1,s_2),g_1\wedge\iota_2(s_2),R_1,(s_1',s_2)):=\lambda_1(s_1,g_1,R_1,s_1')$ and $\gamma((s_1,s_2),g_1\wedge\iota_2(s_2),R_1,(s_1',s_2)):=\gamma_2(s_2)$ (left-sided transitions),
\item $\lambda((s_1,s_2),\iota_1(s_1)\wedge g_2,R_2,(s_1,s_2')):=\lambda_1(s_1)$ and $\gamma((s_1,s_2),\iota_1(s_1)\wedge g_2,R_2,(s_1,s_2')):=\gamma_2(s_2,g_2,R_2,s_2')$ (right-sided transitions).
\end{itemize}
\end{definition}

\begin{figure*}[tbh]
\centering
\begin{subfigure}{0.32\textwidth}
\centering
\includegraphics[scale=0.265]{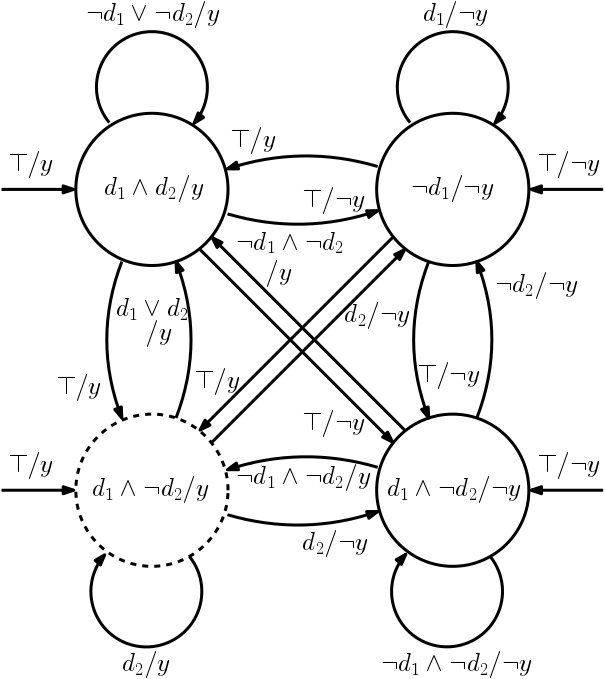}\caption{Timed signal transducer for $U_{(0,\infty)}$}\label{fig:until}
\end{subfigure}
\begin{subfigure}{0.32\textwidth}
\centering
\includegraphics[scale=0.265]{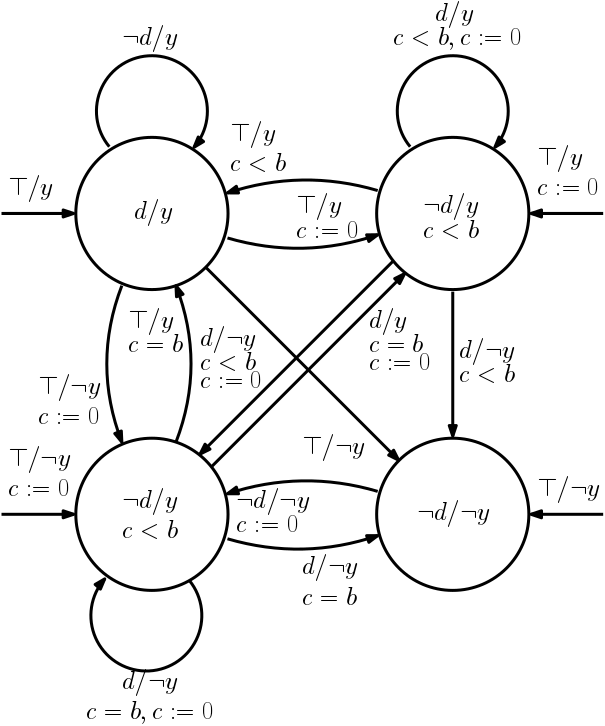}\caption{Timed signal transducer for $F_{(0,b)}$}\label{fig:eventually}
\end{subfigure}
\begin{subfigure}{0.32\textwidth}
\centering
\includegraphics[scale=0.265]{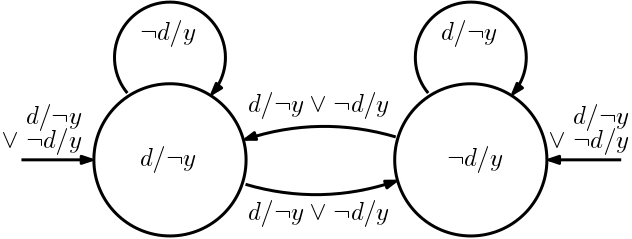}\caption{Timed signal transducer for $\neg$}\label{fig:neg}
\end{subfigure}\\
\begin{subfigure}{0.38\textwidth}
\centering
\includegraphics[scale=0.265]{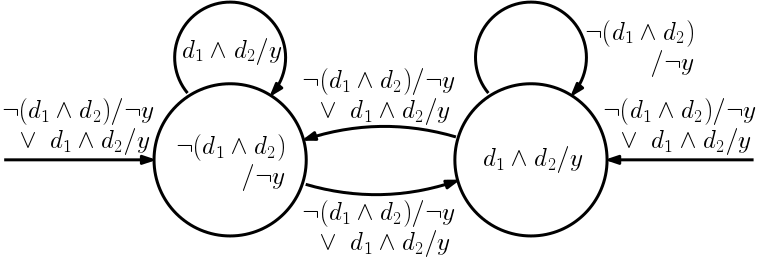}\caption{Timed signal transducer for $\wedge$}\label{fig:conjunction}
\end{subfigure}
\begin{subfigure}{0.58\textwidth}
\centering
\includegraphics[scale=0.265]{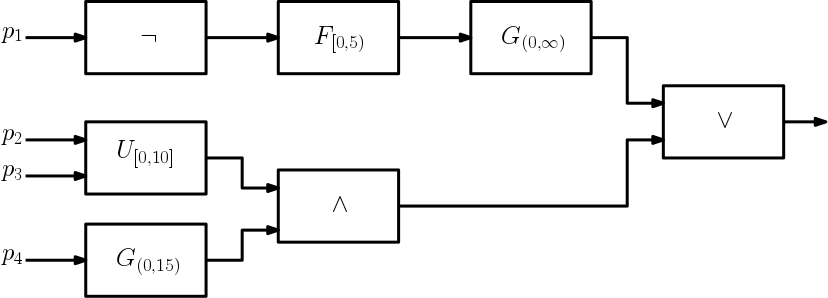}\caption{Formula tree for  $\varphi:=G_{(0,\infty)}F_{[0,5)} \neg p_1 \vee (p_2 U_{[0,10]} p_3 \wedge G_{(0,15)}p_4)$.}\label{fig:formula_tree}
\end{subfigure}
\caption{Timed signal transducers for temporal and Boolean operators and an example formula tree.}
\label{fig:tsd_temporal}
\end{figure*}

We can now summarize the procedure of \cite{ferrere2019real}. First, it shown that every MITL formula $\varphi$ can be rewritten using only temporal operators $U_{(0,\infty)}$ and $F_{(0,b)}$ for rational constants $b$ \cite[Lemmas 4.1 and 4.3]{ferrere2019real}. Second, timed signal transducers for $U_{(0,\infty)}$ and $F_{(0,b)}$ are proposed, see Figs. \ref{fig:until} and \ref{fig:eventually} \cite[Figs. 7 and 11]{ferrere2019real}. Note that all states and transitions except for the state indicated by the dashed line in $U_{(0,\infty)}$ are included in  $\mathcal{F}$. We here further propose timed signal transducers for negations and conjunctions, which are only implicitly mentioned in the proof of \cite[Thm. 6.7]{ferrere2019real}, in Figs. \ref{fig:neg} and \ref{fig:conjunction}. Third, the formula tree of an MITL formula $\varphi$ is constructed as illustrated in Fig. \ref{fig:formula_tree}. Each box in the formula tree represents a timed signal transducer. Boxes not consisting of $\neg$, $\wedge$, $U_{(0,\infty)}$, and $F_{(0,b)}$ can again be rewritten with the results from the first step, i.e., they can be written as a combination of $\neg$, $\wedge$, $U_{(0,\infty)}$, and $F_{(0,b)}$. Fourth, input-output composition and the synchronous product are used to obtain a timed signal transducer $TST_\varphi:=(S,s_0,\Lambda,\Gamma,\boldsymbol{c},\iota,\Delta,\lambda,\gamma, \mathcal{F})$ that has accepting runs over $\boldsymbol{d}$, i.e., $\boldsymbol{d}\in L(TST_\varphi)$, with $\boldsymbol{y}(0)=\top$ (meaning that $\gamma(\delta_0)=y$) if and only if $(\boldsymbol{d},0)\models\varphi$ \cite[Thm. 6.7]{ferrere2019real}. Note that $TST_\varphi$ may have several inputs, but only one output, i.e.,  $\boldsymbol{y}(t)$ is a scalar.

%

\subsection{Problem Formulation}

Consider a dynamical system as given by 
\begin{align}\label{system_noise}
\dot{\boldsymbol{x}}(t)&=f(\boldsymbol{x}(t))+g(\boldsymbol{x}(t))\boldsymbol{u}(t), \; \boldsymbol{x}(0):=\boldsymbol{x}_0
\end{align} 
with  $\boldsymbol{u}(t)\in\mathbb{R}^m$ and $\boldsymbol{x}(t)\in\mathbb{R}^n$. The functions $f:\mathbb{R}^n\to\mathbb{R}^n$ and $g:\mathbb{R}^n\to\mathbb{R}^{n\times m}$ are locally Lipschitz continuous. 
\begin{problem}
Assume that \eqref{system_noise} is subject to an SITL task $\phi$. Derive a control law  $\boldsymbol{u}(\boldsymbol{x},t)$ so that $(\boldsymbol{x},0)\models \phi$ where $\boldsymbol{x}:\mathbb{R}_{\ge 0}\to \mathbb{R}^n$ is the solution to \eqref{system_noise} under $\boldsymbol{u}(\boldsymbol{x},t)$.
\end{problem}
\section{Planning and Control Approach}
\label{sec:strategy}

We first abstract the SITL formula $\phi$ into the MITL formula $\varphi:=Pr(\phi)$. In Section \ref{sec:high_level}, we modify $TST_\varphi$ to account for the error induced by neglecting predicates of $\phi$ in $\varphi$. Based on this modified $TST_\varphi$, denoted by $TST_\phi$, and without considering the dynamics in \eqref{system_noise}, we find high-level plans $d_\mu:\mathbb{R}_{\ge 0}\to BC(M)$ (formally defined below) that characterize all signals $\boldsymbol{x}:\mathbb{R}_{\ge 0}\to\mathbb{R}^n$ such that $(\boldsymbol{x},0)\models\phi$. In Section \ref{sec:abstraction}, we abstract \eqref{system_noise} into a timed signal transducer $TST_S$ that can be used to check if $d_\mu(t)$ can be executed by \eqref{system_noise}. This abstraction is based on the assumption of existing logic-based feedback control laws, such as presented in our works \cite{lindemann2018control,lindemann2019decentralized}. In Section \ref{sec:CEGIS}, we  modify $TST_\phi$ into $TST_\phi^\text{m}$ to ensure that $d_\mu:\mathbb{R}_{\ge 0}\to BC(M)$, now found from $TST_\phi^\text{m}$, can be executed by \eqref{system_noise}.


\subsection{Checking Satisfiability of Signal Interval Temporal Logic}
\label{sec:high_level}
Let $TST_\varphi:=(S,s_{0},\Lambda,\Gamma,\boldsymbol{c},\iota,\Delta,\lambda,\gamma, \mathcal{F})$ be constructed for $\varphi$ according to Section \ref{sec:mitl_to_ta} with $\Lambda:=P$ and $P$ coming from the $Pr(\phi)$ transformation. Since we ultimately aim at satisfying $\phi$, we modify $TST_\varphi$ by the following operations. 
\begin{enumerate}
\item[{[O1]}]  Remove each state $s\in S$ for which there is no $\boldsymbol{x}\in\mathbb{R}^n$ so that $\boldsymbol{x}\models Pr^{-1}(\lambda(s))$. Remove the corresponding $s$ from $\mathcal{F}$. Further remove the corresponding ingoing ($(s',g,R,s)\in\Delta$ for some $s'\in S$) and outgoing ($(s,g,R,s')\in\Delta$ for some $s'\in S$) transitions.
\item[{[O2]}]  Remove each transition $\delta:=(s,g,R,s')\in\Delta$ for which there is no $\boldsymbol{x}\in\mathbb{R}^n$ so that $\boldsymbol{x}\models Pr^{-1}(\lambda(\delta))$. Remove the corresponding $\delta$ from $\mathcal{F}$.
\end{enumerate}

Note that, by employing techniques such as reported in \cite[Ch. 2]{bemporad1999control}, a feasibility problem can be solved to check whether or not there exists $\boldsymbol{x}\in\mathbb{R}^n$ such that $\boldsymbol{x}\models Pr^{-1}(\lambda(s))$ and $\boldsymbol{x}\models Pr^{-1}(\lambda(\delta))$ in $[O1]$ and $[O2]$, respectively. By these operations, we account for predicate dependencies although the planning is performed using propositions. The modified $TST_\varphi$ is denoted by $TST_\phi:=(S^\phi,s_{0},\Lambda,\Gamma,\boldsymbol{c},\iota,\Delta^\phi,\lambda,\gamma, \mathcal{F}^\phi)$ for which naturally $S^\phi\subseteq S$, $\Delta^\phi\subseteq \Delta$, and $\mathcal{F}^\phi\subseteq \mathcal{F}$. The high-level plan synthesis is based on $TST_\phi$ and the fact that we can translate $TST_\phi$, which is in essence a timed automaton \cite{alur1994theory} when removing the output labels, to a region automaton $RA(TST_\phi)$; $RA(TST_\phi)$ can be used to check emptiness of $TST_\phi$, i.e., to analyze reachability properties of $TST_\phi$. Since $TST_\phi$ has invariants on states $\iota(s)$ and guards $g$ included in transitions  $(s,g,R,s')\in\Delta^\phi$, we have to modify the algorithm presented in \cite{alur1994theory}. Note that the timed automaton in \cite{alur1994theory} only possesses guards and labels on transitions, while the timed automaton in \cite{alur1996benefits} only has invariants and labels on states so that we here have a hybrid of these two. Similarly to \cite{alur1996benefits}, we associate a transition relation $\Rightarrow$ over the extended state space $S^\phi\times \mathbb{R}_{\ge 0}^O$ as follows: $(s,\boldsymbol{c},\delta)\Rightarrow (s',\boldsymbol{c}')$ if and only if there exist $t',t''\in\mathbb{R}_{\ge 0}$ and $\delta:=(s,g,R,s')\in\Delta^\phi$ so that
\begin{itemize}
\item for all $\tau\in[0,t')$, $\boldsymbol{c}+\tau\models\iota(s)$,
\item for all $\tau\in(t',t'+t'']$, $R(\boldsymbol{c}+t')+\tau\models\iota(s')$,
\item it holds that $\boldsymbol{c}':=R(\boldsymbol{c}+t')+t''$ and  $\boldsymbol{c}+t'\models g$,
\end{itemize}
i.e., a combination of continuous evolution and discrete transition. Reachability properties of the infinite state transition system $(S^\phi\times \mathbb{R}_{\ge 0}^O,\Rightarrow)$ can now be analyzed by its finite state region automaton $RA(TST_\phi)$ that relies on a bisimulation relation $\sim\subseteq \mathbb{R}^O_{\ge 0}\times \mathbb{R}^O_{\ge 0}$ resulting in clock regions. In fact, a clock region is an equivalence class induced by $\sim$.  Details are omitted and the reader is referred to \cite{alur1994theory} for details. Let $\alpha$ and $\alpha'$ be clock regions and assume $\boldsymbol{c}\in\alpha$ and $\boldsymbol{c}'\in\alpha'$. If $(s,\boldsymbol{c},\delta)\Rightarrow (s',\boldsymbol{c}')$ and $\boldsymbol{c}\sim \bar{\boldsymbol{c}}$ for some $\bar{\boldsymbol{c}}$, it then holds that there is a $\bar{\boldsymbol{c}}'$ with $\boldsymbol{c}'\sim \bar{\boldsymbol{c}}'$ so that  $(s,\bar{\boldsymbol{c}},\delta)\Rightarrow (s',\bar{\boldsymbol{c}}')$.

\begin{definition}
The region automaton $RA(TST_\phi):=(Q,q_0,\Rightarrow_R,\mathcal{F}_R)$ is the quotient system of $(S^\phi\times \mathbb{R}_{\ge 0}^O,\Rightarrow)$ using clock regions as equivalence classes and  defined as:
\begin{itemize}
\item The states are $(s,\alpha)$ where $s\in S^\phi$ and $\alpha\in A$ where $A$ is the set of all clock regions so that $Q:=S^\phi\times A$.
\item The initial states are $q_0:=(s_0,\alpha_0)\in Q$ where $\alpha_0$ is the clock region corresponding to $\boldsymbol{c}(0)$.
\item There is a transition $(s,\alpha,\delta)\Rightarrow_R (s',\alpha')$ if and only if there is a transition $(s,\boldsymbol{c},\delta)\Rightarrow (s',\boldsymbol{c}')$ for $\boldsymbol{c}\in\alpha$ and $\boldsymbol{c}'\in\alpha'$.
\item $(s,\alpha)\in \mathcal{F}_R$ if $s\in \mathcal{F}^\phi$.
\end{itemize}
\end{definition}

Using standard graph search techniques such as the memory efficient variant of the nested depth first search \cite{courcoubetis1992memory}, here adapted to deal with the generalized B\"uchi acceptance condition as in \cite{tauriainen2006nested}, we may obtain, if existent, sequences $\bar{s}=((s_0,\alpha_0),(s_1,\alpha_1),\hdots)$ with $(s_j,\alpha_j,\delta_j)\Rightarrow_R(s_{j+1},\alpha_{j+1})$ for each $j\in \mathbb{N}$ satisfying the generalized B\"uchi acceptance condition $\mathcal{F}_R$. In particular, $\bar{s}:=(\bar{s}_p,\bar{s}_p^\omega)$ consists of a prefix of length $p+1$ and a suffix of length $s$, here denoted by $\bar{s}_p:=((s_0,\alpha_0), \hdots, (s_p,\alpha_p))$ and  $\bar{s}_s:=((s_{p+1},\alpha_{p+1}),\hdots,(s_{p+s},\alpha_{p+s})) $. Furthermore, we require that $\gamma(\delta_0)=y$ to indicate that we want $(\boldsymbol{d},0)\models \varphi$, opposed to $\gamma(\delta_0)=\neg y$ indicating $(\boldsymbol{d},0)\models \neg \varphi$. What remains to be done is to add timings $\bar{\tau}:=(\bar{\tau}_p,\bar{\tau}_s^\omega)$ to $\bar{s}$ with, similarly to $\bar{s}_p$ and $\bar{s}_p$, $\bar{\tau}_p:=(\tau_0:=0,\hdots,\tau_p)$ and $\bar{\tau}_s:=(\tau_{p+1},\hdots,\tau_{p+s})$ where $\tau_j\in\mathbb{R}_{>0}$  for $j \ge 1$ corresponds to the occurence of $\delta_j$, which happens $\tau_j$ time units after the occurence of $\delta_{j-1}$. The proof of \cite[Lemma~4.13]{alur1994theory} proposes a method to find timings for a simple acceptance condition, i.e., only requiring $\bar{\tau}_p$, while we deal with a generalized B\"uchi acceptance condition for which we present a solution in Section \ref{sec:CEGIS} and assume, for now, that  $\bar{\tau}$ has been obtained.  Such $\bar{s}$ with $\bar{\tau}$ can be associated, by denoting $T_j:=\sum_{k=0}^j \tau_j$, with a high-level plan (later interpreted as $d_\mu(t)$) as
\begin{align}\label{eq:plan1133}
d_p(t):=\begin{cases}
\lambda(\delta_j) &\text{if } t= T_j\\ 
\lambda(s_j) &\text{if }  T_j < t <T_{j+1}
\end{cases}
\end{align}

\begin{lemma}\label{lem:1}
Assume a signal $\boldsymbol{d}:\mathbb{R}_{\ge 0}\to\mathbb{B}^{|P|}$. There is an accepting run of $TST_\phi$ over $\boldsymbol{d}(t)$ and $(\boldsymbol{d},0)\models \varphi$ if only if there exists a plan $d_p(t)$ so that  $\boldsymbol{d}(t) \models d_p(t)$ for all $t\in\mathbb{R}_{\ge 0}$.

\begin{proof}
$\Rightarrow$: Departing from $TST_\phi$, the infinite state transition system $(S\times \mathbb{R}_{\ge 0}^O,\Rightarrow)$ has, by construction, the same reachable set as $TST_\phi$, i.e, the same reachable configurations $(s_0,\boldsymbol{c}(0)),(s_0,R(\boldsymbol{c}(0))), (s_1,R(\boldsymbol{c}(0))+\tau_1),\hdots$. Since $\sim$ is a bisimulation relation, reachability properties of $TST_\phi$ can then equivalently be analyzed by considering the finite state transition system $RA(TST_\phi)$ \cite[Lemma 4.13]{alur1994theory}. If there hence exists an accepting run of $TST_\phi$ over $\boldsymbol{d}(t)$ and $(\boldsymbol{d},0)\models \varphi$, i.e., $\gamma(\delta_0)=y$, the plan $d_p(t)$ can be constructed as described above by obtaining $\bar{s}$ and $\bar{\tau}$ directly from the accepting run of $TST_\phi$ over $\boldsymbol{d}(t)$. It will, by construction, hold that $\boldsymbol{d}(t)\models d_p(t)$ for all $t\in\mathbb{R}_{\ge 0}$.

$\Leftarrow$: Finding accepting runs $\bar{s}$ of $RA(TST_\phi)$ using nested depth first search algorithms, and including suitable timings $\bar{\tau}$, ensures that $TST_\phi$ has an accepting run for an input signal $\boldsymbol{d}(t)\models d_p(t)$ for all $t\in\mathbb{R}_{\ge 0}$. Removing states and transitions from $TST_\varphi$ according to operations $[O1]$ and $[O2]$ resulting in $TST_\phi$ only removes behavior from $TST_\varphi$ (not adding additional behavior), i.e., $L(TST_\phi) \subseteq L(TST_\varphi)$, so that, by \cite[Thm. 6.7]{ferrere2019real}, an accepting run of $TST_\phi$ over $\boldsymbol{d}(t)$ inducing $\boldsymbol{y}(0)=\top$ results in $(\boldsymbol{d},0)\models \varphi$. 
\end{proof}
\end{lemma}

Note that there may exist an accepting run of $TST_\varphi$ over $\boldsymbol{d}(t)$ so that $(\boldsymbol{d},0)\models \varphi$, while there exists no accepting run of $TST_\phi$ over $\boldsymbol{d}(t)$ due to operations $[O1]$ and $[O2]$. We can now associate $d_\mu:\mathbb{R}_{\ge 0}\to BC(M)$ with $d_p(t)$ by letting $d_\mu(t):=Pr^{-1}(d_p(t))$ and, based on $\phi$, state under which conditions $d_p(t)$ exists.

\begin{lemma}\label{lem:2}
There exists a plan $d_p(t)$ (and hence a plan $d_\mu(t)$) if and only if $\phi$ is satisfiable.

\begin{proof}
Recall that $TST_\varphi$  has an accepting run over $\boldsymbol{d}(t)$ with $\boldsymbol{y}(0)=\top$ if and only if $(\boldsymbol{d},0)\models\varphi$. Operations $[O1]$ and $[O2]$ remove all states and transitions from $TST_\varphi$ that are infeasible, i.e., for which there exists no $\boldsymbol{x}\in\mathbb{R}^n$ such that $\boldsymbol{x}\models Pr^{-1}(\lambda(s))$ and $\boldsymbol{x}\models Pr^{-1}(\lambda(\delta))$, respectively. Since the only difference between the semantics of $\phi$ and $\varphi$ is the difference in the semantics of $\mu_i$ and $p_i$, respectively, the following holds:

$\Rightarrow$: The existence of a plan $d_p(t)$ implies, by Lemma \ref{lem:1}, that any signal $\boldsymbol{d}:\mathbb{R}_{\ge 0}\to\mathbb{B}^{|P|}$ with $\boldsymbol{d}(t)\in d_p(t)$ for all $t\in\mathbb{R}_{\ge 0}$ is such that $(\boldsymbol{d},0)\models \varphi$. It follows that there exists a signal $\boldsymbol{x}:\mathbb{R}_{\ge 0}\to\mathbb{R}^n$ with $\boldsymbol{x}(t)\models d_\mu(t)$ for all $t\in\mathbb{R}_{\ge 0}$  implying that $(\boldsymbol{x},0)\models \phi$, i.e., $\phi$ is satisfiable.

$\Leftarrow$: If $\phi$ is satisfiable, it means that there exists a signal $\boldsymbol{x}:\mathbb{R}_{\ge 0}\to\mathbb{R}^n$ such that $(\boldsymbol{x},0)\models \phi$. Associated with $\boldsymbol{x}(t)$, define the signal $\boldsymbol{d}(t):=\begin{bmatrix}
h_1^\top(\boldsymbol{x}(t)) & \hdots & h_M^\top(\boldsymbol{x}(t))
\end{bmatrix}^T$ that is such that $(\boldsymbol{d},0)\models\varphi$ and where $h_i^\top(\boldsymbol{x}):=\top$ if $h_i(\boldsymbol{x})\ge 0$ and $h_i^\top(\boldsymbol{x}):=\bot$ otherwise. Note that $h_i(\boldsymbol{x})$ is the predicate function associated with $\mu_i$. It follows that $\boldsymbol{d}$ induces an accepting run of $TST_\phi$  over $\boldsymbol{d}$ so that, by Lemma \ref{lem:1}, it follows that there hence exists a plan $d_p(t)$. 
\end{proof}
\end{lemma}

\begin{theorem}\label{thm:main}
If a signal $\boldsymbol{x}:\mathbb{R}_{\ge 0}\to \mathbb{R}^n$ is such $\boldsymbol{x}(t)\models d_\mu(t)$ for all $t\in\mathbb{R}_{\ge 0}$, then it follows that $(\boldsymbol{x},0)\models \phi$.

\begin{proof}
Follows from the proof of Lemma \ref{lem:2}.
\end{proof}
\end{theorem}

\subsection{Timed Abstraction of the Dynamical Control System}
\label{sec:abstraction}

 
We abstract the system in \eqref{system_noise} into a timed signal transducer $TST_S:=(\tilde{S},\tilde{S}_0,\tilde{\Lambda},\tilde{c},\tilde{\Delta},\tilde{\lambda})$. Note the absence of output labels, invariants, and a B\"uchi acceptance condition, and that $\tilde{c}$ is a scalar. The previous notation of a plan $d_\mu(t)$ will allow $TST_S$ to be an acceptor or a refuser of such a high-level plan $d_\mu(t)$, i.e., $TST_S$ will indicate if the dynamics in \eqref{system_noise} in conjunction with a feedback control law $\boldsymbol{u}(\boldsymbol{x},t)$ can execute the required motion according to $d_\mu(t)$ to satisfy $\phi$. The transition relation $\tilde{\Delta}$ is now based on the ability of the system to switch in finite time, by means of a feedback control law $\boldsymbol{u}_{\tilde{\delta}}(\boldsymbol{x},t)$  between elements in $E:=Pr^{-1}(BC(TST_\phi))\subseteq BC(\tilde{\Lambda})$ where $\tilde{\Lambda}:=M$ and $BC(TST_\phi):=\{z\in BC(P)| \exists s \in S\cup \Delta, \lambda(s)=z \}$.  It is assumed that a library of such logic-based feedback control laws $\boldsymbol{u}_{\tilde{\delta}}(\boldsymbol{x},t)$ is available, e.g., as in \cite{lindemann2018control,lindemann2019decentralized}. Assume that $|\tilde{S}|= |E|$ and let $\tilde{\lambda}:\tilde{S}\to E$ where, for $\tilde{s}',\tilde{s}''\in \tilde{S}$ with $\tilde{s}'\neq \tilde{s}''$, it holds that $\tilde{\lambda}(\tilde{s}')\neq\tilde{\lambda}(\tilde{s}'')$ so that each state is uniquely labelled by $\tilde{\lambda}$, i.e., each state indicates exactly one Boolean formula from $E$. Note that $TST_\phi$ and $TST_S$ now align in a way that will allow to avoid the state space explosion when forming a product automaton between them. A transition from $\tilde{s}$ to $\tilde{s}'$ is indicated by $(\tilde{s},\tilde{g},0, \tilde{s}')\in \tilde{\Delta}$ where $\tilde{g}$ is a guard that depends on \eqref{system_noise}. In particular, we assume that $\tilde{g}$ encodes intervals of the form $(C',C'')$, $[C',C'')$, $(C',C'']$, $[C',C'']$, or conjunctions of them, where $C',C''\in\mathbb{Q}_{\ge 0}$ with $C'\le C''$. There exists a transition $\tilde{\delta}:=(\tilde{s},\tilde{g},0, \tilde{s}')\in \tilde{\Delta}$ if, for all $\tau>0$ with $\tau\models \tilde{g}$ and for all $\boldsymbol{x}_0\in\mathbb{R}^n$ with $\boldsymbol{x}_0\models \tilde{\lambda}(\tilde{s})$, there exists a control law $\boldsymbol{u}_{\tilde{\delta}}(\boldsymbol{x},t)$ so that the solution $\boldsymbol{x}(t)$ to \eqref{system_noise} is such that: 
\begin{itemize}
\item either, for all $t\in [0,\tau)$, $\boldsymbol{x}(t)\models \tilde{\lambda}(\tilde{s})$  and  $\boldsymbol{x}(\tau)\models \tilde{\lambda}(\tilde{s}')$
\item or, for all $t\in [0,\tau]$, $\boldsymbol{x}(t)\models \tilde{\lambda}(\tilde{s})$ and there exists $\tau'>\tau$ such that, for all $t\in(\tau,\tau']$, $\boldsymbol{x}(\tau')\models \tilde{\lambda}(\tilde{s}')$.
\end{itemize}

We define $\tilde{\lambda}(\tilde{\delta}):=\tilde{\lambda}(\tilde{s}')$ in the former and $\tilde{\lambda}(\tilde{\delta}):=\tilde{\lambda}(\tilde{s})$ in the latter case. Note that $\boldsymbol{u}_{\tilde{\delta}}(\boldsymbol{x},t)$, achieving such a transition, has to ensure invariance and finite-time reachability properties. If these control laws are according to \cite{lindemann2018control,lindemann2019decentralized}, we emphasize that it is ensured that the solution $\boldsymbol{x}(t)$ to \eqref{system_noise} is defined for all $t\in\mathbb{R}_{\ge 0}$; $\tilde{S}_0$ is here a set and consists of all element $\tilde{s}_0\in \tilde{S}$ such that $\boldsymbol{x}_0\models\tilde{\lambda}(\tilde{s}_0)$. We now define a run of $TST_S$ slightly different compared to a run of $TST_\phi$. A run of $TST_S$ over the input signal $d_\mu:\mathbb{R}_{\ge 0}\to BC(M)$ again consists of an alternation of time and discrete steps $(\tilde{s}_0,0)\xrightarrow{\tilde{\delta}_0}(\tilde{s}_1,0)\xrightarrow{\tau_1}(\tilde{s}_1,\tau_1)\xrightarrow{\tilde{\delta}_1} \hdots$. A time step of duration $\tau$ is denoted by $(\tilde{s},0)\xrightarrow{\tau}(\tilde{s},\tau)$ with $d_\mu(t+t')=\tilde{\lambda}(\tilde{s})$ for each $t'\in (0,\tau)$. A discrete step at time $t$ is denoted by $(\tilde{s},\tilde{c}(t))\xrightarrow{\tilde{\delta}}(\tilde{s}',0)$ for some transition $\tilde{\delta}=(\tilde{s},\tilde{g},0,\tilde{s}')\in\tilde{\Delta}$ such that $\tilde{c}(t)\models \tilde{g}$ and for which $\boldsymbol{x}\models\tilde{\lambda}(\tilde{\delta})$  implies that $\boldsymbol{x}\models d_\mu(t)$. If $d_\mu(t)$ does not result in a run of $TST_S$ over $d_\mu(t)$, then it can  be concluded that \eqref{system_noise} can not execute $d_\mu(t)$. Otherwise, i.e., $d_\mu(t)$ results in a run of $TST_S$ over $d_\mu(t)$, we define the control law $\boldsymbol{u}(\boldsymbol{x},t)$ based on the plan $d_\mu(t)$ and the run of $TST_S$ over $d_\mu(t)$. Recall the definition of $T_j$ and let $\boldsymbol{u}(\boldsymbol{x},t):=\boldsymbol{u}_{\tilde{\delta}_{1}}(\boldsymbol{x},t)$ for all $t\in[0,T_1)$, $\boldsymbol{u}(\boldsymbol{x},t):=\boldsymbol{u}_{\tilde{\delta}_{j+1}}(\boldsymbol{x},t-T_j)$ for all $t\in(T_j,T_{j+1})$ with $j\ge 2$, and $\boldsymbol{u}(\boldsymbol{x},T_j):=\boldsymbol{u}_{\tilde{\delta}_{j+1}}(\boldsymbol{x},0)$ (or $\boldsymbol{u}(\boldsymbol{x},T_j):=\boldsymbol{u}_{\tilde{\delta}_{j}}(\boldsymbol{x},\tau_j)$) for $j\ge 2$ if $\boldsymbol{x}\in\mathbb{R}^n$ with $\boldsymbol{x}\models\tilde{\lambda}(\tilde{s}_{j+1})$ (or $\boldsymbol{x}\models\tilde{\lambda}(\tilde{s}_{j})$)  implies that $\boldsymbol{x}\models d_\mu(T_j)$.

\begin{theorem}\label{thm:2}
If $d_\mu(t)$ results in a run of $TST_S$ over $d_\mu(t)$, then applying $\boldsymbol{u}(\boldsymbol{x},t)$ to \eqref{system_noise} results in $(\boldsymbol{x},0)\models \phi$.  

\begin{proof}
If $d_\mu(t)$ results in a run of $TST_S$ over $d_\mu(t)$, applying $\boldsymbol{u}(\boldsymbol{x},t)$ to \eqref{system_noise} results in $\boldsymbol{x}(t)\models d_\mu(t)$ for all $t\in \mathbb{R}_{\ge 0}$ due to the way transitions $\tilde{\delta}$ in $TST_S$ are defined. According to Theorem \ref{thm:main}, we can  infer that $(\boldsymbol{x},0)\models \phi$.
\end{proof}
\end{theorem}


\subsection{Plan Synthesis for Signal Interval Temporal Logic}
\label{sec:CEGIS}

Sections \ref{sec:high_level} and \ref{sec:abstraction} present a way to synthesize $d_\mu(t)$ that can be checked against $TST_S$ as in Theorem \ref{thm:2}. It may, however, occur that $d_\mu(t)$ does not result in a run of $TST_S$ due the system in \eqref{system_noise} being unable to follow $d_\mu(t)$. We propose a complete algorithm that avoids a state space explosion that is typically the outcome of forming automata products. This follows since the input label of each state or transition in $TST_\phi$ corresponds to one state in $TST_S$, i.e., $TST_\phi$ and $TST_S$ align in a way, so that $TST_\phi^\text{m}$ (defined below and corresponding to the product of $TST_\phi$ and $TST_S$) has no more states than $TST_\phi$.
\begin{remark}
The usefulness of avoiding such state explosion is illustrated as follows. If $\varphi$ is build from three elementary signal transducers,  e.g., one until and two eventually operators as in Figs. \ref{fig:until} and \ref{fig:eventually}, $TST_\phi$ will have $4^3=64$ states in the worst case (depending on the operations $[O1]$ and $[O2]$). Assuming a discrete abstraction $DA$ of \eqref{system_noise}, such as a weighted transition system \cite{nikou2016cooperative}, with $100$ states, e.g. corresponding to a discretization of $\mathbb{R}^n$, the product of $TST_\phi$ and $DA$ may contain up to $6400$ states. The situation gets even worse when forming the region automaton which induces $\mathcal{O}(|S|\cdot 2^{C_\text{max}})$ states where $S$ are the states of the product automaton and $C_\text{max}$ is the maximum clock constant contained in $S$ \cite[Thm.~4.16]{alur1994theory}. 
\end{remark}

Our approach relies on two facts: 1) the removal of states and edges, as presented in $[O1]$ and $[O2]$ and continued below, resulting in $TST_\phi^\text{m}$ and 2) constraining guards $g$ of transitions in $TST_\phi$ so  that it is possible to determine timings $\bar{\tau}$, if possible, for $\bar{s}$ that result in $d_\mu(t)$ being a run of $TST_S$. We modify $TST_\phi$ to account for $TST_S$ as follows.
\begin{enumerate}
\item[{[O3]}] Remove each transition $\delta:=(s,g,R,s')\in \Delta^\phi$ for which there exists no transition $\tilde{\delta}:=(\tilde{s},\tilde{g},0,\tilde{s}')\in \tilde{\Delta}$  with $\lambda(s)=Pr(\tilde{\lambda}(\tilde{s}))$, $\lambda(s')=Pr(\tilde{\lambda}(\tilde{s}'))$, and for which $\boldsymbol{x}\models\tilde{\lambda}(\tilde{\delta})$ implies $\boldsymbol{x}\models Pr^{-1}(\lambda(\delta))$. Remove the corresponding $\delta$ from $\mathcal{F}^\phi$.
\item[{[O4]}]  Remove each $\delta_0:=(s_0,g,R,s')\in \Delta$ with $\boldsymbol{x}_0\not\models Pr^{-1}(\lambda(s'))$. Remove the corresponding $\delta_0$ from $\mathcal{F}^\phi$.
\end{enumerate}

Denote the obtained sets by $S^\text{m}$, $\Delta^\text{m}$, and $\mathcal{F}^\text{m}$ for which  $S^\text{m}\subseteq S^\phi$, $\Delta^\text{m}\subseteq \Delta^\phi$, and $\mathcal{F}^\text{m}\subseteq \mathcal{F}^\phi$. Operation $[O3]$ removes transitions in $TST_\phi$ for which there exists no corresponding transition in $TST_S$, while operation $[O4]$ takes care of the initial position $\boldsymbol{x}_0$. We further take care of the timings including an additional clock into $TST_\phi$. Therefore, let $\boldsymbol{c}^\text{m}:=\begin{bmatrix}
\boldsymbol{c}^T & \tilde{c}\end{bmatrix}^T$ and perform the final operation.
\begin{enumerate}
\item[{[O5]}] For each transition $\delta^\text{m}:=(s,g,R,s')\in\Delta^\text{m}$ let $g^\text{m}=g \wedge \tilde{g}$ where $\tilde{\delta}:=(\tilde{s},\tilde{g},0,\tilde{s}')\in\tilde{\Delta}$ with $\lambda(s)=Pr(\tilde{\lambda}(\tilde{s}))$, $\lambda(s')=Pr(\tilde{\lambda}(\tilde{s}'))$, and for which $\boldsymbol{x}\models\tilde{\lambda}(\tilde{\delta})$ implies $\boldsymbol{x}\models Pr^{-1}(\lambda(\delta))$. Replace $g$ and $R$ in $\delta^\text{m}$ with $g^\text{m}$ and $R^\text{m}$, respectively, where $R^\text{m}$ is obtained in an obvious manner. 
\end{enumerate}

We emphasize that adding $\tilde{c}$ and $\tilde{g}$ is crucial to ensure correctness. Let the modified timed signal transducer be denoted by $TST^\text{m}_\phi:=(S^\text{m},s_{0},\Lambda,\Gamma,\boldsymbol{c}^\text{m},\iota,\Delta^\text{m},\lambda,\gamma, \mathcal{F}^\text{m})$ and note that $L(TST^\text{m}_\phi)\subseteq L(TST_\phi)\subseteq L(TST_\varphi)$.
\begin{remark}
The operations $[O3]$-$[O5]$ result in the timed signal transducer $TST^\text{m}_\phi$ that restricts the behavior of $TST_\phi$ exactly to the behavior allowed by $TST_S$ and corresponds hence to a product automaton without exhibiting an exponential state space explosion.
\end{remark} 

We next explain how to find an accepting plan $d_p(t)$ from $TST^\text{m}_\phi$. Similar to Section \ref{sec:high_level}, we find $\bar{s}$ by a nested depth first search now performed on $RA(TST^\text{m}_\phi)$. Recall that the nested depth first search provides $\bar{s}$ that consist of a prefix $\bar{s}_p$ and a suffix $\bar{s}_s$. For such a sequence $\bar{s}$, we now determine if suitable timings $\bar{\tau}_p$ and $\bar{\tau}_s$ can be found so that the resulting $d_\mu(t)$ is accepted by $TST_S$. For ease of reading, we use  $TST_\phi$ and $TST_S$ (and not $TST_\phi^\text{m}$) in the remainder. Note also that the guards $g$ and invariants $\iota(s)$ in $TST_\phi$ are always conjunctions of the form $c_o< C_o$ or $c_o=C_o$ for clocks $o\in\{1,\hdots,O\}$ where $C_o\in\mathbb{Q}_{\ge 0}$, while the guards $\tilde{g}$ in $TST_S$ are, by assumption, always of the form $(C',C'')$, $[C',C'')$, $(C',C'']$, and $[C',C'']$, or conjunctions of them.

\textbf{Prefix ($\tau_0$ - $\tau_{p}$):} Note first that $\tau_0:=0$.  For  $\tau_j$ with $j\in\{1,\hdots, p\}$, the transitions $(s_j,\alpha_j,\delta_j)\Rightarrow_R(s_{j+1},\alpha_{j+1})$ have to be considered where $\delta_j:=(s_j,g_j,R_j,s_{j+1})\in\Delta^\phi$. For each such transition $\delta_j$, let $\tilde{g}_j$ be the corresponding guard in $TST_S$ in accordance with operation $[O3]$, i.e., $\tilde{\delta}_j:=(\tilde{s}_j,\tilde{g}_j,0,\tilde{s}_{j+1})\in \tilde{\Delta}$ of $TST_S$ with $\lambda(s_j)=Pr(\tilde{\lambda}(\tilde{s}_j))$, $\lambda(s_{j+1})=Pr(\tilde{\lambda}(\tilde{s}_{j+1}))$,  and for which $\boldsymbol{x}\models\tilde{\lambda}(\tilde{\delta}_j)$ implies $\boldsymbol{x}\models Pr^{-1}(\lambda(\delta_j))$. Let $N_{j,o}$ with $0\le N_{j,o}\le j$ denote the number of preceding transitions in which the clock $c_o$ was not reset, i.e., $R_k(c_o)=c_o$ for all  $k\in\{j-N_{j,o},\hdots,j-1\}$ and $R_{j-N_{j,o}-1}(c_o)=0$ if $j-N_{j,o}>0$. With this definition, let us further define $T_{j,o}:=\sum_{k=j-N_{j,o}}^{j-1}\tau_k$ if $N_{j,o}>0$ and $T_{j,o}:=0$ if $N_{j,o}=0$.  We next consider four cases for determining the timings $\bar{\tau}_p$. 

Case 1) If there exists $o\in \{1,\hdots,O\}$ so that $\boldsymbol{c}(\tau_j)\models g_j$ only if $c_o(\tau_j)=C_o$, then it has to hold that 
\begin{align}\label{eq:1}
\tau_j\in \{C_o-T_{j,o}\} \cap \tilde{g}_j.
\end{align}

Otherwise, i.e., $\boldsymbol{c}(\tau_j)\models g_j$ does not imply that there exists  $o\in \{1,\hdots,O\}$ so that $c_o(\tau_j)=C_o$, partition $\{1,\hdots,O\}$ as $\{1,\hdots,O\}:=O_{R,j}\cup O_{NR,j}$ so that  $R_j(c_o)=0$ for all $o\in O_{R,j}$, while $R_j(c_o)=c_o$ for all $o\in O_{NR,j}$. Let $\bar{c}_{j,o}$ and $\underline{c}_{j,o}$ be the $o$th elements of $\bar{\boldsymbol{c}}_{j}:=\text{argsup}_{\boldsymbol{c}\in\alpha_{j+1}} \|\boldsymbol{c}\|$ and $\underline{\boldsymbol{c}}_{j}:=\text{arginf}_{\boldsymbol{c}\in\alpha_{j+1}} \|\boldsymbol{c}\|$, respectively. 
 
Case 2) If not Case 1  and, for some $o\in O_{NR,j}$, we have $\underline{c}_{j,o}=\bar{c}_{j,o}$, then, for $o\in O_{NR,j}$, it has to hold that
\begin{align}\label{eq:2}
\tau_j\in \{\bar{c}_{j,o}-T_{j,o}\} \cap \tilde{g}_j.
\end{align}

Case 3) If not Cases 1 and 2, then, for $o\in O_{NR,j}$, let $\delta_{j,o}':=\underline{c}_{j,o}-T_{j,o}$ and $\delta_{j,o}'':= \bar{c}_{j,o}-T_{j,o}$. 

Case 4) If not Cases 1 and 2, then, for $o\in O_{R,j}$, let $\delta_{j,o}':=0$ and $\delta_{j,o}'':=C_o- T_{j,o}$. 

For Cases 3 and 4, we then require that 
\begin{align}\label{eq:34}
\tau_j\in\big(\max_{o\in\{1,\hdots,O\}}\delta_{j,o}'+\epsilon,\min_{o\in\{1,\hdots,O\}}\delta_{j,o}''\big) \cap \tilde{g}_j
\end{align} where $\epsilon>0$  avoids Zeno signals and guarantees progressive runs \cite{alur1994theory}. We see that \eqref{eq:1}-\eqref{eq:34} are constrained, in a similar way, by $\tilde{g}_j$ which exactly corresponds to operation $[O5]$.

\textbf{Suffix ($\tau_{p+1}$ - $\tau_{p+s}$):} The suffix can be found in a similar way as the prefix, i.e., considering Cases 1-4. We only need to add a lasso shape condition. In other words, we find $\tau_{p+1}$ until $\tau_{p+s}$ as described in Steps 1-4, but now additionally requiring that, for $o\in O_{NR,p+1}$, 
\begin{align}\label{eq:lasso}
T_{p,o}=T_{p+s,o}.
\end{align}

To obtain $\bar{\tau}$ and check if there exists a $\bar{\tau}$ corresponding to $\bar{s}$, consider the following optimization problem.
\begin{subequations}\label{eq:feas_problem}
\begin{align}
&\text{arg min}_{\bar{\tau}\in\mathbb{R}_{\ge 0}^{p+s+1},\epsilon\in\mathbb{R}_{> 0}} \epsilon\\
\text{s.t. } &\tau_0:=0,  \epsilon>0\\
& \tau_j \text{ for } j\in \{1,\hdots,p-1\} \text{ according to \eqref{eq:1}-\eqref{eq:34}}\\
& T_{p,o}=T_{p+s,o} \text{ for } o\in O_{NR,p+1} \text{ according to \eqref{eq:lasso}}.
\end{align}
\end{subequations}

\begin{corollary}
The optimization problem in \eqref{eq:feas_problem} is a linear and hence convex optimization problem.

\begin{proof}
Note that Cases 1-4 can not happen simultaneously. The constraint in \eqref{eq:1} can be written as $\tau_j=C_o-T_{j,o}$ and $\tau_j\models \tilde{g}_j$. The latter constraint can be written into separate linear constraints using the constants $C_j'$ and $C_j''$ associated with $\tilde{g}_j$. Note that $T_{j,o}$ is a linear combination of $\tau_j$'s. Hence \eqref{eq:1} is a linear constraint in $\bar{\tau}$. It is straighforward to show the same for \eqref{eq:2} and \eqref{eq:lasso}. The constraint \eqref{eq:34} can be written into constraints $\tau_j> \delta_{j,o}'+\epsilon$ and $\tau_j<\delta_{j,o}''$ for each $o\in\{1,\hdots,O\}$ where $\delta_{j,o}'$ and $\delta_{j,o}''$ are again linear in $\bar{\tau}$. It can also be seen that \eqref{eq:34} is linear in $\epsilon$. The optimization problem in \eqref{eq:feas_problem} is hence linear and thus convex.
\end{proof}
\end{corollary}

Note in particular that \eqref{eq:feas_problem} is always feasible.  With $\bar{s}$ and $\bar{\tau}$ as obtained above, we can then synthesize $d_\mu(t)$ as in \eqref{eq:plan1133}.

\begin{theorem}\label{thm:3}
The proposed method is sound. Given a timed abstraction $TST_S$, the proposed method is also complete.

\begin{proof}
Regarding soundness. If the nested depth first search finds $\bar{s}$ and $\bar{\tau}$ is obtained from \eqref{eq:feas_problem}, then $d_\mu(t)$ results in a run of $TST_S$ over $d_\mu(t)$ due to operations $[O3]$-$[O5]$ performed on $TST_\phi$ resulting in $TST_\phi^\text{m}$. Applying $\boldsymbol{u}(\boldsymbol{x},t)$ to \eqref{system_noise} then results in $(\boldsymbol{x},0)\models \phi$ according to Theorem \ref{thm:2}. Note that Theorems \ref{thm:main} and \ref{thm:2} hold even when $d_\mu(t)$ is obtained from $TST_\phi^\text{m}$ instead of $TST_\phi$ since $L(TST_\phi^\text{m})\subseteq L(TST_\phi)$.
 
Regarding completeness. The proof of Lemma \ref{lem:2} shows completeness for plans found from $TST_\phi$, but without considering $TST_S$; $TST_\phi^\text{m}$  restricts the language of $TST_\phi$ by considering $TST_S$ and only removing behavior that $TST_S$ can not execute. Hence we can find a plan $d_\mu(t)$ from $TST_\phi^\text{m}$ if there exists a plan $d_\mu(t)$ that is accepted by $TST_S$.  
\end{proof}
\end{theorem}

Note that Theorem \ref{thm:3} guarantees completeness on the planning level, i.e., when given an abstraction $TST_S$.

%
%
%
%
%
%

\section{Simulations}
\label{sec:simulations}

\begin{figure}
\centering
\input{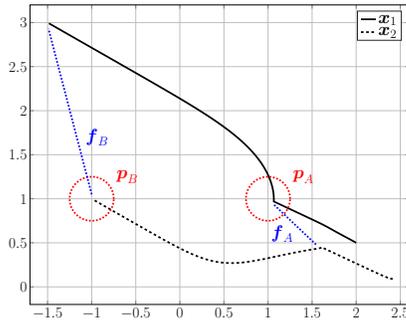}
\caption{Execution of the plan $d_\mu(t)$ by applying $\boldsymbol{u}(x,t)$. The timings $\bar{\tau}$ are respected since $\boldsymbol{x}(1)\models \mu_2$ and $\boldsymbol{x}(2)\models \mu_3$ (indicated by the red dotted circles $\boldsymbol{p}_A$ and $\boldsymbol{p}_B$).}
\label{fig:all}
\end{figure}

We consider an academic example that is easy to follow, yet rich enough to illustrate the theoretical findings of this paper. Consider a system consisting of $\boldsymbol{x}:=\begin{bmatrix}
\boldsymbol{x}_1^T & \boldsymbol{x}_2^T
\end{bmatrix}^T\in\mathbb{R}^{4}$, e.g., a system consisting of two robots. The SITL formula is $\phi:=(\mu_1 U_{(0,\infty)} \mu_2) \wedge F_{(0,3)} \mu_3 \wedge F_{(0,3)} \mu_4$ with predicate functions $h_1(\boldsymbol{x}):= \epsilon-\|\boldsymbol{x}_1-\boldsymbol{x}_2 -\boldsymbol{f}_A\|$, $h_2(\boldsymbol{x}):= \epsilon-\|\boldsymbol{x}_1-\boldsymbol{p}_A\|$, $h_3(\boldsymbol{x}):= \epsilon-\|\boldsymbol{x}_2 -\boldsymbol{p}_B\|$, and $h_4(\boldsymbol{x}):= \epsilon-\|\boldsymbol{x}_1-\boldsymbol{x}_2 -\boldsymbol{f}_B\|$ and it holds that $\boldsymbol{x}_0\models \mu_1$, while $\boldsymbol{x}_0\not\models \mu_2$, $\boldsymbol{x}_0\not\models \mu_3$, and $\boldsymbol{x}_0\not\models \mu_4$ (important for operation $[O4]$). Let $\epsilon:=0.25$ and $\boldsymbol{f}_A:=\begin{bmatrix}
-0.5 & 0.5
\end{bmatrix}^T$ and $\boldsymbol{f}_B:=\begin{bmatrix}
-0.5 & 2
\end{bmatrix}^T$ so that $\mu_1$ and $\mu_4$ encode formations between the robots. Let further $\boldsymbol{p}_A:=\begin{bmatrix}
1 & 1
\end{bmatrix}^T$ and $\boldsymbol{p}_B:=\begin{bmatrix}
-1 & 1
\end{bmatrix}^T$ so that $\mu_2$ and $\mu_3$ encode reachability specifications of robots $1$ and $2$, respectively. Note that there is no $\boldsymbol{x}\in\mathbb{R}^4$ so that $\boldsymbol{x}\models (\mu_2 \wedge \mu_3) \vee (\mu_1 \wedge \mu_4)$ (important for operations $[O1]$ and $[O2]$). The corresponding MITL formula $\varphi:=Pr^{-1}(\phi)=(p_1 U_{(0,\infty)} p_2) \wedge F_{(0,3)} p_3 \wedge F_{(0,1)} p_4$ was translated to $TST_\varphi$ resulting in $65$ states. We assume the dynamics $\dot{\boldsymbol{x}}=f(\boldsymbol{x})+\boldsymbol{u}$ where $f(\boldsymbol{x})$ may be unknown and consider, for instance, control laws as derived in \cite{lindemann2018control}. These control laws can achieve invariance and finite time reachability specifications. In other words, there exists control laws $\boldsymbol{u}_{\tilde{\delta}}(\boldsymbol{x},t)$ that can satisfy STL formulas such as $G_{[0,b)} \mu_\text{inv}\wedge F_{[b]} \mu_\text{reach}$  in case that the predicate function associated with $\mu_\text{inv}$ and $\mu_\text{reach}$ are concave and satisfiable, which is the case for conjunctions of $\mu_1$-$\mu_4$. We also assume, for simplicity, that each possible transition $\tilde{\delta}$ can be made within $1$-$4$ time units, i.e., $C':=1$ and $C'':=4$ so that $b\in [C',C'']$ (important for operation $[O5]$); $TST_\varphi$ was then transformed into $TST_\phi$ which was again transformed into $TST_\phi^\text{m}$ by performing operations $[O1]$-$[O5]$. Based on this, $RA(TST_\phi^\text{m})$ was obtained with, in total, $2723$ states. To illustrate that our method deals with spatiotemporal specifications, we compare the nested depth first search of $RA(TST_\phi^\text{m})$ with a nested depth first search performed on $RA(TST_\varphi)$. For $RA(TST_\varphi)$, a sequence $\bar{s}^\varphi:=((s_{0}^\varphi,\alpha_{0}^\varphi),(s_{1}^\varphi,\alpha_{1}^\varphi),\hdots)$ is obtained with $\lambda(s_{0}^\varphi)=p_1\wedge \neg p_2\wedge \neg p_3 \wedge \neg p_4$ and $\lambda(s_{1}^\varphi)=p_1\wedge  p_2\wedge  p_3 \wedge  p_4$ for which there exist no $\boldsymbol{x}\in\mathbb{R}^n$ so that $\boldsymbol{x}\models Pr^{-1}(\lambda(s_{1}^\varphi))$. Our method, however, finds a sequence $\bar{s}:=((s_{0},\alpha_{0}),(s_{1},\alpha_{1}),\hdots)$ from $RA(TST_\phi^\text{m})$ with $\lambda(s_0)=p_1\wedge \neg p_2\wedge \neg p_3 \wedge \neg p_4$, $\lambda(s_1)=p_1\wedge  p_2\wedge  \neg p_3 \wedge  \neg p_4$, and $\lambda(s_2)=\neg p_1 \wedge  p_3 \wedge  p_4$ accounting for the spatial properties induced by the predicates, i.e., for each $\lambda(s_0)$, $\lambda(s_1)$, and $\lambda(s_1)$ there exists $\boldsymbol{x}\in\mathbb{R}^n$ so that $\boldsymbol{x}\models Pr^{-1}(\lambda(s_{0}))$, $\boldsymbol{x}\models Pr^{-1}(\lambda(s_{1}))$, and $\boldsymbol{x}\models Pr^{-1}(\lambda(s_{2}))$, respectively. A timing sequence $\bar{\tau}$ is obtained with $\bar{\tau}:=(0,1,1,1,\hdots)$ defining the plan $d_\mu(t)$ that can be implemented as stated in Theorem \ref{thm:3}, resulting in $(\boldsymbol{x},0)\models \phi$ as shown in Fig. \ref{fig:all}.

\section{Conclusion}
\label{sec:conclusion}

This paper presents an efficient automata-based planning and control framework for spatio-temporal logics, here in particular signal interval temporal logic. Results from automata-based verification for metric interval temporal logic have been leveraged to account for the spatial properties induced by the signal interval temporal logic specification at hand. Furthemore, the state explosion, typically induced by forming a product automaton between the specification automaton and an abstraction of the system, is avoided. For future work, we will consider the  robust semantics as well as uncontrollable events within the planning framework.

\bibliographystyle{IEEEtran}
\bibliography{literature}

\addtolength{\textheight}{-12cm}   

\end{document}